\documentclass[prl,twocolumn,showpacs,superscriptaddress]{revtex4}

\usepackage{graphicx}
\usepackage{color}

\begin{document}

\title{High-pressure study of X-ray diffuse scattering in ferroelectric perovskites.}
             
\author{Sylvain Ravy}
\affiliation{Synchrotron SOLEIL, L'Orme des merisiers, Saint-Aubin BP 48, 
91192 Gif-sur-Yvette cedex, France}
\author{Jean-Paul Iti\'e}
\affiliation{Synchrotron SOLEIL, L'Orme des merisiers, Saint-Aubin BP 48, 
91192 Gif-sur-Yvette cedex, France}
\affiliation{Physique des Milieux Denses, IMPMC, CNRS UMR 7590, Universit\'e P. et M. Curie,
140 rue de Lourmel, 75015 Paris, France.}
\author{Alain Polian}
\affiliation{Physique des Milieux Denses, IMPMC, CNRS UMR 7590, Universit\'e P. et M. Curie,
140 rue de Lourmel, 75015 Paris, France.}
\author{Michael Hanfland}
\affiliation{European Synchrotron Radiation Facility, 6 rue Jules Horowitz,
Bo\^ite postale 220, 38043 Grenoble Cedex, France}

\begin{abstract}
We present a high-pressure x-ray diffuse scattering study of the ABO$_3$ ferroelectric perovskites BaTiO$_3$ and KNbO$_3$.
The well-known diffuse lines are observed in all the phases studied.
In KNbO$_3$, we show that the lines are present up to 21.8 GPa, with constant width and a slightly decreasing intensity.
At variance,  the intensity of the diffuse lines observed in the cubic phase of BaTiO$_3$ linearly decreases to zero at $\sim 11$ GPa.
These results are discussed with respect to x-ray absorption measurements, which leads to the conclusion that the diffuse lines are only observed when the B atom is off the center of the oxygen tetrahedron.
The role of such disorder on the ferroelectric instability of perovskites is discussed.
\end{abstract}

\pacs{61.10.-i, 61.50.Ks, 77.84.Dy}

\maketitle

Because of the simplicity of their structure and their actual or potential applications, perovskites ABO$_3$ and their derivatives are amongst the most studied compounds in solid state science.
Perovskites stabilize a large variety of states such as superconductivity, ferroelectricity (FE) and multiferroism, but despite decades of studies, the mechanisms of these transitions are still a matter of debate \cite{Ferroelectricity}.
This is the case of the oxygen perovskites such as SrTiO$_3$ and KTaO$_3$, which only exhibit incipient FE, while PbTiO$_3$, BaTiO$_3$ and KNbO$_3$ stabilize this instability in well defined series of FE transitions.

Upon heating at ambient pressure, barium titanate and potassium niobate undergo the same famous sequence of rhomboedric-orthorhombic-tetragonal-cubic (R-O-T-C) first-order phase transitions \cite{Hewat,Comes0}. 
During this sequence, the polarization changes direction from the [111], to the [110] and [100] pseudo-cubic directions, before vanishing above the Curie temperature $T_C$ in the cubic phase.
The mechanism of this sequence of phase transition has been often discussed from two limiting cases: 
the soft mode - or displacive - theory, in which a transverse optical (TO) mode frequency gradually soften when approaching the phase transition \cite{Cochran}, and an order-disorder scenario \cite{Aubry}, in which some atoms lying on symmetry-equivalent sites, order on preferential ones.
In fact, current theories favor an intermediate situation, in which soft phonons and disorder play a role \cite{Girshberg,Pirc}. 

Pressure has proven to be an essential parameter to study phase transitions in perovskites \cite{Samara}.
Because applying pressure results in a symmetrization of chemical bonds, drastic effects on order-disorder phase transitions are expected.
Indeed, the same sequence of phase transitions is observed as a function of pressure in KNbO$_3$ and BaTiO$_3$ \cite{Pruzan}.
At 300 K, FE is lost at 10 GPa (2 GPa) and 
is no more observed at low temperature above $\sim26$~GPa ($\sim6.5$ GPa) in KNbO$_3$ (BaTiO$_3$).
The existence of the O-T transition under pressure has been questioned in KNbO$_3$ \cite{Kobayashi}, while in other studies its transition pressure was found at 6 GPa \cite{Pruzan0} or 8.5 GPa \cite{Pruzan}.
In this letter, we will show that high pressure x-ray diffuse scattering measurements give a unique opportunity to study the transitions and the FE instability as a function of the nature and the position of the B atom.

\vspace{-2mm}
\begin{figure}[htp]
\centering
\includegraphics[width=0.46\textwidth]{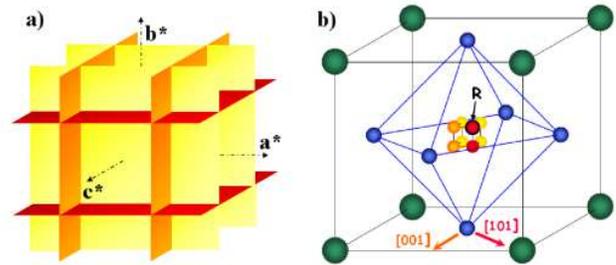}
\caption[Diffuse lines]{(Color online) a) Schematic representation of the diffuse sheets in the reciprocal space.
The color scheme corresponds to the sites occupancy shown in b).
b) Perovkites pseudo-cubic unit cell and the 8-sites of the Ti/Nb atoms.
The R, red, red-orange, and red-orange-yellow sites are occupied in the R-, O-, T-, and C-phases respectively.
For the sake of consistency with previous studies, we take the {\bf b} axis vertical, and {\bf c} along the x-ray beam.
}
\label{DL}
\vspace{-3mm}
\end{figure}


The fluctuations leading to the FE state are complex, as demonstrated by the numerous and apparently conflicting experimental results gathered since decades.
After the first observation of diffuse lines (DL) by electron \cite{Honjo} and X-ray diffraction \cite{Harada1} in barium titanate, Com\`es {\it et al.} \cite{Comes0} performed a thourough study of the DL in potassium niobate, and showed that they are due to reciprocal diffuse \{100\} {\it sheets} (see Fig. \ref{DL}a)) intersecting the Ewald sphere. 
There are three sets of sheets in the C phase, two sets in the T phase ((010) in red and (100) in orange), one in the O phase (010) and no DL in the low temperature R phase.
Measured far from the Bragg positions, the intensity and the width of the lines are constant in a given phase \cite{footnote1}.

\begin{figure}[ht]
\centering
\includegraphics[height=0.2\textwidth]{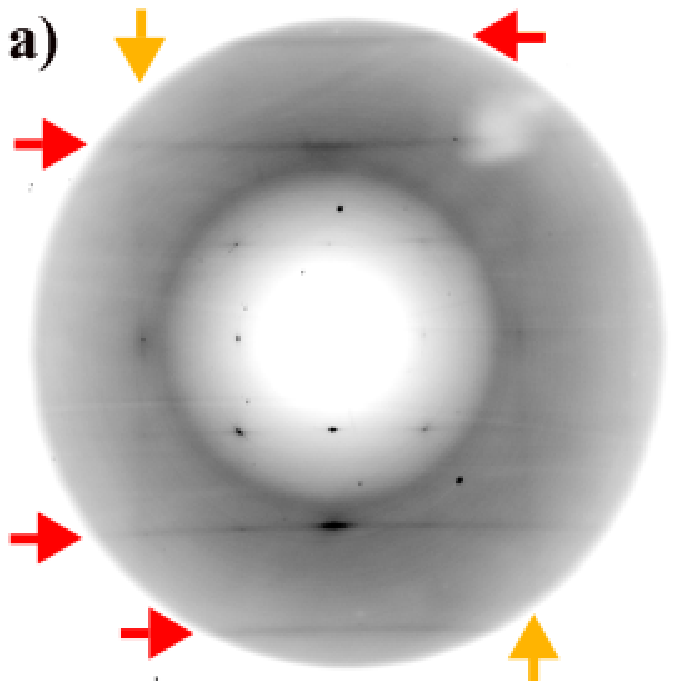} 
\includegraphics[height=0.2\textwidth]{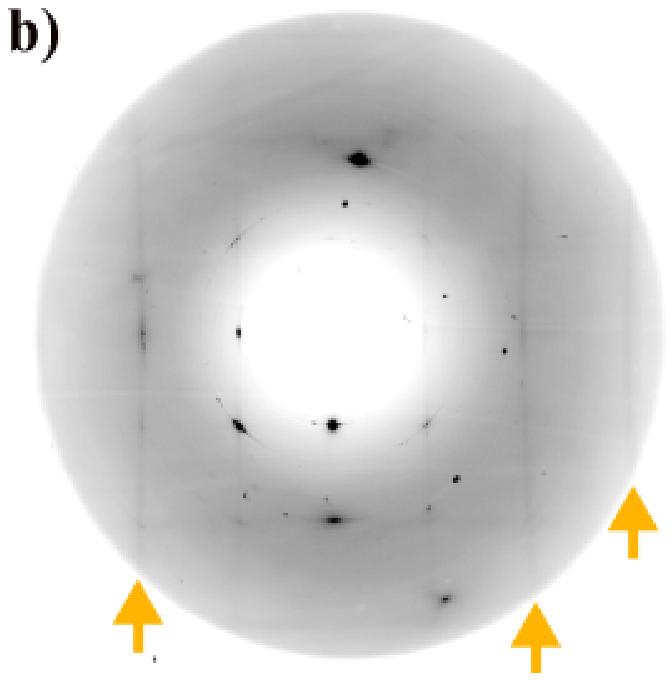}
\includegraphics[height=0.2\textwidth]{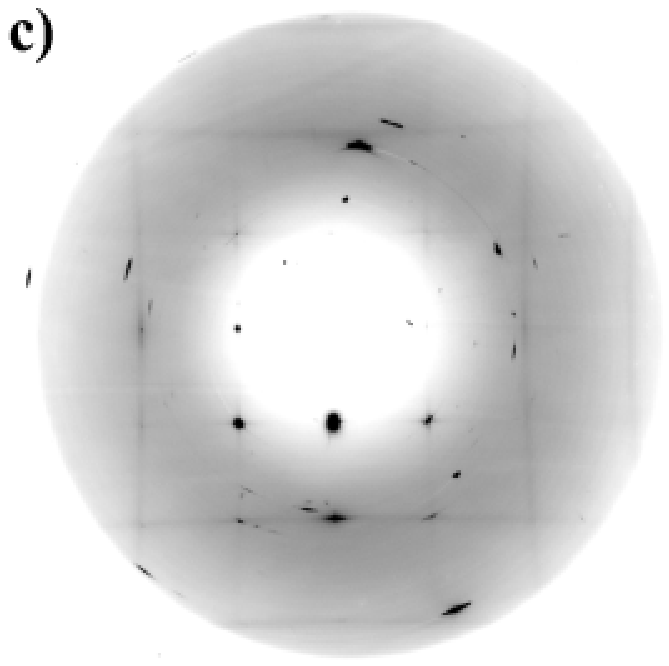}
\includegraphics[height=0.2\textwidth]{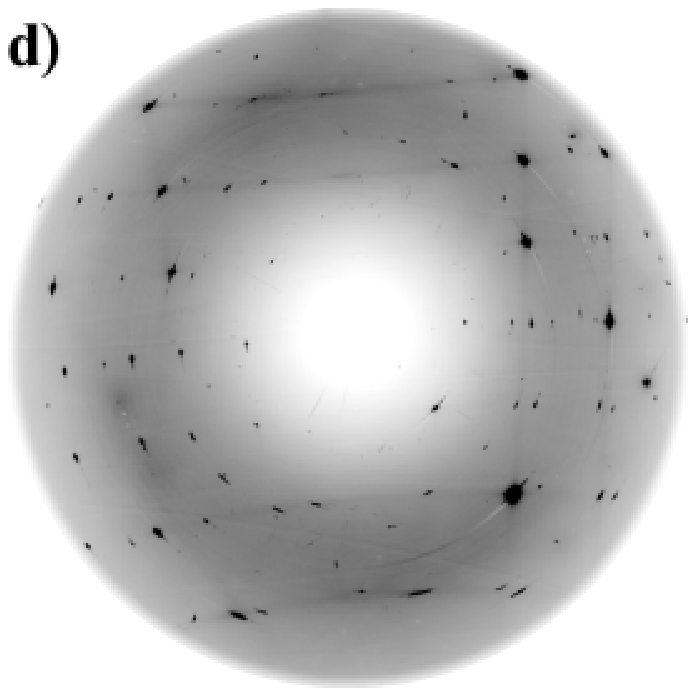}\\
\caption{(Color online) Scattering patterns of KNbO$_3$ (\#1) at a) 4.3 GPa, b) 6.3 GPa and c) 12.1 GPa.
Red and orange arrows point to (010) and (100) diffuse lines respectively.
The background and the diffuse ring in a) are due to the scattering by liquid Ne.
d) Diffraction pattern of KNbO$_3$ (\#2) at 21.8 GPa.}
\label{KNbO3}
\vspace{-3mm}
\end{figure}

Com\`es {\it et al.} interpreted the DL from the so-called "8-sites" model \cite{Takahashi} in which the B atom is locally displaced from the center of the unit cell along the $\langle$111$\rangle$ directions (see Fig. \ref{DL}b)).
This model was actually very efficient in explaining the sequence of phase transitions by the successive ordering of the B atoms on 4, 2 and 1 sites in the T, O and R phases respectively.
The DL were shown to be due to independent linear correlations of each components of the Nb (Ti) displacement.

To explain these results, H\"uller \cite{Huller} proposed an alternative model based on the presence of highly anisotropic TO soft mode.
In fact, inelastic neutron scattering (INS) revealed a more complex, and phase-dependant origin of the DL.
In O-KNbO$_3$, Currat {\it et al.} \cite{Currat,Currat2} clearly evidenced i) a highly anisotropic transverse acoustic (TA) branch with a flat dispersion at $\sim $1.6~THz in the diffuse scattering plane, and ii) a well defined soft TO mode with similar anisotropy.
This gave a clear support to the soft-mode interpretion, at least in the O phase.
At variance, in C-KNbO$_3$ \cite{Nunes} and in C-BaTiO$_3$ \cite{Harada2} an anisotropic and highly {\it overdamped} TO mode was observed, which was rather interpreted in the order-disorder framework.

Inelastic light scattering results \cite{Fontana3,Sokoloff,Dougherty} complexified this picture, by evidencing a "central peak" owing to a relaxational mode, in O-, T-KNbO$_3$, T- and C-BaTiO$_3$.
This mode was interpreted from the 8-sites model as due to thermally activated jumps between {\it unsymmetrical} \cite{Sokoloff,Dougherty} positions of the Ti or Nb atoms.

\begin{figure}[th]
\centering
\includegraphics[width=0.2\textwidth]{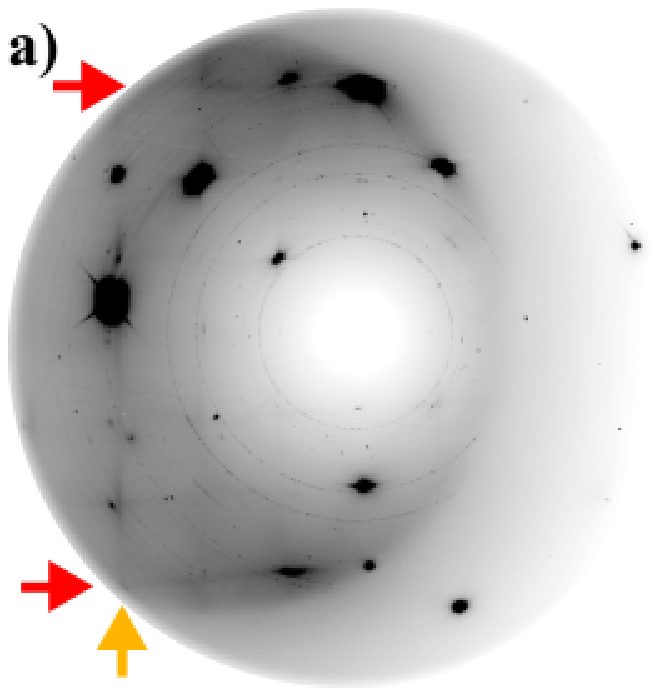} 
\includegraphics[width=0.2\textwidth]{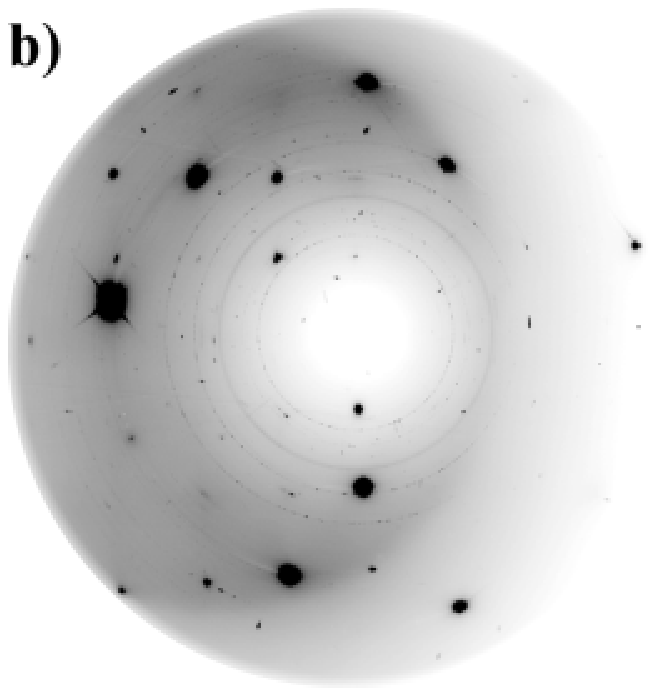}\\
\caption{(Color online) Scattering patterns of BaTiO$_3$ at a) 2.6 GPa and b) 11.3 GPa.
Red and orange arrows point to (010) and (100) diffuse lines respectively.
The contrast has been increased, which makes the diffuse lines more visible and the Bragg spots
more intense than in Fig. \ref{KNbO3}.}
\label{BaTisample3}
\end{figure}

Results from local probes techniques like EPR \cite{Muller1}, NMR \cite{Zalar} or XAFS
\cite{Kim,Ravel,Frenkel}, also support the 8-sites model.
In particular, it was found by EXAFS in all the phases of KNbO$_3$,\cite{Kim} that Nb is shifted by $u\sim0.2$~\AA\ with respect to the oxygen octahedron, in the [111] direction.
This value is consistent with the displacements measured in the ordered phases \cite{Hewat}.
In the case of BaTiO$_3$, a pre-edge peak at the Ti K-edge, whose intensity is proportional to $\langle u^2 \rangle$ \cite{Vedrinskii}, was studied as a function of the x-ray polarization, which lead to the same conclusion \cite{Ravel}.
Thus, whatever the actual type of dynamics, these XAFS results show that in the paraelectric phase, the Ti(Nb) atoms spend more time at the [111] sites than at any other sites of the unit cell, including its center.

Most interesting is the recent evidence by XAFS measurements that in BaTiO$_3$ the Ti atom moves and locks on to the center of the oxygen octahedron above 10 GPa \cite{Itie}, at variance with KNbO$_3$, where the Nb atom does not reach the central position up to 15.8 GPa \cite{Frenkel}.
This stimulated us to carry out high-pressure X-ray diffuse scattering experiments on KNbO$_3$ and BaTiO$_3$.


The experiments have been performed at the ID09A beam line of the European Synchrotron Radiation Facility (ESRF), using a membrane diamond anvil cell (MDAC) \cite{LeToullec}.
The 30$\times$30 $\mu$m$^2$ 30 keV beam was vertically focused by a spherical mirror and horizontally by a bent Silicon (111) monochromator.
All the diffraction patterns presented here were collected by a Mar345 image-plate detector, using 2$^\circ$-oscillations of the crystal about the vertical axis on a -15$^\circ$ to 15$^\circ$ angular domain.

A first KNbO$_3$ single crystal was loaded in the hole of a stainless steel gasket (initial diameter $\Phi$=150 $\mu$m, initial thickness $e$=52 $\mu$m), with Ne as pressure-transmitting medium, and a ruby sphere as pressure probe.
Typical diffraction patterns are shown on Fig. \ref{KNbO3} a), b) and c).
In this MDAC, the maximum measurable scattering vector was $q\sim 5$~\AA$^{-1}$.
As this crystal broke at about 13 GPa, a second single crystal was mounted in a He-loaded MDAC (maximum $q\sim 6.2$~\AA$^{-1}$), with a stainless steel gasket ($\Phi$=250 $\mu$m, $e$=80 $\mu$m).
This mounting allowed us to keep the crystal integrity up to 21.8 GPa (Fig. \ref{KNbO3}d)).
\begin{figure}[th]
\centering
\includegraphics[width=0.45\textwidth]{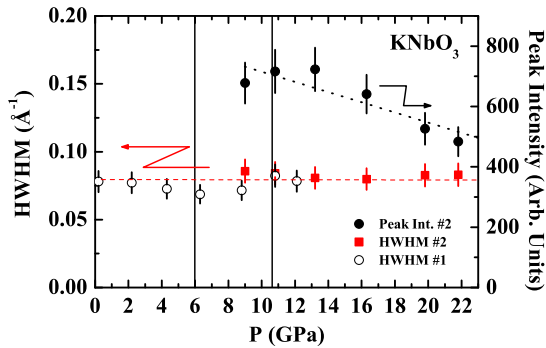}
\vspace{-5mm}
\caption{(Color online) Pressure dependence of the peak intensity (right scale), and the HWHM (left scale) of the diffuse lines of KNbO$_3$ (sample 1(2), open(full) symbols).
Lines are guides for the eye.
Vertical lines indicate the O-T and T-C transition pressures, according to this study}
\label{IvsTKNbO3}
\vspace{-5mm}
\end{figure}
The BaTiO$_3$ single crystal was loaded in the hole of a Rh gasket ($\Phi$=120 $\mu$m, $e$=70 $\mu$m) with methanol-ethanol-water (16:3:1), in the first type of MDAC.

The diffraction patterns of KNbO$_3$ displayed in Fig. \ref{KNbO3} clearly show the presence of \{001\} DL in all the phases.
The signal-to-noise ratio for the DL was found to be about 10\%, probably because of the diamond Compton scattering. 
In the O-phase, (010) DL are visible up to $k=3$, with similar characteristics as first reported in \cite{Comes0}: absence of $k=0$ lines, and Half-Width at Half-Maximum (HWHM) equal to $\sim0.08$ \AA$^{-1}$ (1/20 of the BZ).
By increasing the pressure below $\sim6$ GPa (Fig. \ref{KNbO3}a)), very weak (100) DL can be detected.
Between $\sim6$ GPa and $\sim10$ GPa, (Fig. \ref{KNbO3}b)), the second set of (100) DL becomes visible, and appears to be stronger than the (010) set.
This result was observed on the two samples.
Above 10 GPa (Fig. \ref{KNbO3}c) and d)), the two sets of (010) and (100) DL are present with the same intensity, indicating the stabilization of the C phase, consistently with the results of Refs. \cite{Pruzan0,Kobayashi,Pruzan}.
Note that in the C-phase the set of (001) DL should be visible, but a look to Fig. 6a) of Ref. \cite{Comes0} makes it clear that these DL are weaker, and only apparent at the $h,k=4$ DL level.
These levels were unreachable in our geometry.
The observation of a change above 6 GPa and 10 GPa in both samples, clearly indicates phase transitions, that we assign to O-T and T-C, in agreement with \cite{Pruzan0}.
However, the unexpected relative intensity of the DL may indicate an unusual orientation of the tetragonal domains. 

\begin{figure}[ht]
\centering
\includegraphics[width=0.450\textwidth]{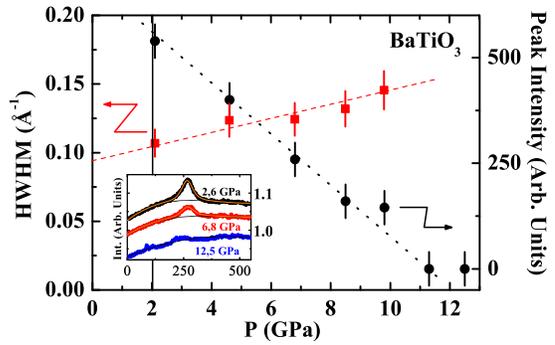}
\vspace{-5mm}
\caption{(Color online) Pressure dependence of the peak intensity (right scale), and the HWHM (left scale) of the diffuse lines of BaTiO$_3$.
Lines are guides for the eye.
The vertical line indicates the T-C transition pressure.
Insert: profiles of the diffuse lines at 2.6, 6.8 and 12.5 GPa.}
\label{IvsTBaTiO3}
\vspace{-5mm}
\end{figure}

The behavior of the HWHM and the intensity of the $k=3$ DL is indicated in Fig. \ref{IvsTKNbO3}.
Only the intensity measured on sample 2 are shown, due to the difficulty of normalization between different samples.
These graphs show that i) the HWHM $\sim0.08$ \AA$^{-1}$ is constant throughout the explored pressure range, ii) the intensity slowly decreases at high pressure.

Typical diffraction patterns in BaTiO$_3$ are displayed on Fig. \ref{BaTisample3}.
As already noted in previous studies \cite{Comes0}, the DL are weaker in BaTiO$_3$ than in KNbO$_3$.
Morover, they appear to be less intense near the ZB boundary than in the ZB center \cite{Takesue}.
The pressure dependence of the intensity and the HWHMs of the DL (measured at the ($\overline{3}$,$\overline{1.8}$,0) reduced scattering vector) are shown in Fig. \ref{IvsTBaTiO3}.
The intensity decreases linearly, and becomes undetectable at $\sim11$~GPa, while the HWHM  is slightly increasing.


Let us recall that for an atom of scattering factor $f$ displaced by $u$, the total diffuse scattering intensity is proportional to $f^2 \langle u^2\rangle$.
$\langle u^2\rangle$ being similar in both compounds \cite{Frenkel,Ravel}, the intensity of the KNbO$_3$ DL is clearly due to the large atomic number of Nb.
Moreover, INS results have shown that despite similar anisotropy, the TA mode is less dispersive in KNbO$_3$ than in BaTiO$_3$ (see Fig. 1 in \cite{Currat2} and Fig. 5 in \cite{Harada2}). 
This explains the uniformity of the KNbO$_3$ DL.
In BaTiO$_3$, the contribution of the TA mode in weaker near the ZB boundary.

Most interesting is to compare x-ray and XAFS results.
In KNbO$_3$, the weakening of the DL under pressure corresponds to the decrease of $u$ measured by EXAFS \cite{Frenkel}.
Even more striking is the strong correlation in BaTiO$_3$, between the vanishing of the DL at $\sim11$ GPa and the saturation of the intensity of the pre-edge peak at the same pressure \cite{Itie}.
As both the DL intensity and the XANES pre-edge peak are proportional to $\langle u^2\rangle$, they have the same pressure dependence (nearly linear).

Our main results is to show unambiguously that the intensity of the DL is correlated with the off-centering displacements of the B atoms, as measured by XAFS.
Consequently, the peculiar dynamics associated to the FE instability, and giving rise to the DL, results from the $\langle$111$\rangle$ displacements of the B atom.
The behavior observed in FE PbTiO$_3$, in which Ti is found to be displaced in the $\langle$100$\rangle$ direction \cite{Ravel}, while no DL were observed by x-ray scattering \cite{Chapman}, reinforces the conclusion that the DL are due to the $\langle$111$\rangle$ local displacements.

These results suggest drastic changes in the dynamics under pressure, especially in C-BaTiO$_3$.
Firstly, the contribution of the TA mode near the BZ boundary, and of the TO mode at the zone center should decrease under pressure. 
We expect the TA and the TO modes to strongly harden up to $\sim 11$ GPa. 
Secondly, INS experiments have shown that increasing the disorder makes the TO mode more and more damped. 
We thus expect that the centering of Ti under pressure, which decreases the disorder, makes the TO mode behave in a more harmonic way.
Finally, we suggest that the relaxational mode as observed by hyperRaman \cite{Fontana3}, and probably by NMR \cite{Zalar,Stern}, whose existence is due to jumps of Ti on unequivalent sites, should disappear above $\sim11$ GPa.
The present observation of a weak (100) set of DL in O-KNbO$_3$ at 4.3 GPa (also noted in \cite{Comes0} below the O-T transition temperature), could be the scattering signature of this mode.

Considering that the FE state is not observed above 6.5 GPa in BaTiO$_3$, it is likely that BaTiO$_3$ is an incipient FE between 6.5 and 11 GPa, and KNbO$_3$ above $\sim26$ GPa.
It is thus tempting to conclude that a large enough off-centering is necessary to stabilize the FE state.
From the value obtained at 6.5 GPa, this threshold value can be estimated from XANES \cite{Ravel,Itie} to be $u_{Ti}=0.1$ \AA\ for BaTiO$_3$, and 0.08 \AA\ for KNbO$_3$ (extrapolated at 26 GPa).
Remarkably, no disorder is found in the incipient FEs KTaO$_3$ \cite{Terauchi}, and in SrTiO$_3$ recent XAFS studies conclude to 0.08 \AA\ off-centering of Ti \cite{EXAFSSr}, a value slightly smaller than $u_{Ti}$.
On the theoretical side, it is noteworthy that recent {\it ab initio} calculations have shown that the FE instability is suppressed at $\sim20$ GPa \cite{Bousquet} or at $\sim10$ GPa \cite{Kornev}.

In conclusion, we have shown that high-pressure XAFS and x-ray scattering experiments allow one to disentangle the role of the lattice dynamics and the local order of the B atoms in the ferroelectric instabilities of perovskites. 
In particular, our results demonstrate that the B atom disorder enhances the FE instability, and support the idea that it is an essential ingredient for the stabilization of the FE state.  

We thank R. Com\`es, R. Currat, and B. Dorner for useful discussions.


\begin{thebibliography}{1}
\bibitem{footnote1} In Ref. \cite{Comes0}, the intensity was measured at the (0,2,0.75) reduced scattering vector
\end{thebibliography}
\end{document}